\documentclass{my_nature}
\usepackage{amssymb,graphicx,url}
\title{Kinematic detection of a planet carving a gap in a protoplanetary disc}

\author{C. Pinte$^{1,2}$, G.~van der Plas$^{2}$, F . M\'enard$^{2}$, D. J.~Price$^{1}$,  V.~Christiaens$^{1}$, T.~Hill$^{3}$, D.~Mentiplay$^{1}$, C.~Ginski$^{4}$, E. Choquet$^{5}$, Y.~Boehler$^{2}$, G.Duch\^ene$^{6,2}$, S.~Perez$^{7}$, S.~Casassus$^{8}$
}

\begin{document}

\maketitle
\begin{affiliations}
\item Monash Centre for Astrophysics (MoCA) and School of Physics and Astronomy, Monash University, Clayton Vic 3800, Australia,
\item Univ. Grenoble Alpes, CNRS, IPAG, F-38000 Grenoble, France
\item Atacama Large Millimeter/Submillimeter Array, Joint ALMA Observatory, Alonso de C\'ordova 3107, Vitacura 763-0355, Santiago, Chile
\item Anton Pannekoek Institute for Astronomy, University of Amsterdam, Science Park 904,1098XH Amsterdam, The Netherlands
\item Aix Marseille Univ, CNRS, CNES, LAM, Marseille, France
\item Astronomy Department, University of California, Berkeley, CA 94720-3411, USA
\item Universidad de Santiago de Chile, Av. Libertador Bernardo O’Higgins 3363, Estación Central, Santiago
\item Departamento de Astronom\'ia, Universidad de Chile, Casilla 36-D, Santiago, Chile
\end{affiliations}

\begin{abstract}
We still do not understand how planets form, or why extra-solar planetary systems are so different from our own solar system.
But the last few years have dramatically changed our view of the discs of gas and dust around young stars.
Observations with the Atacama Large Millimeter/submillimeter Array (ALMA) and extreme adaptive-optics systems have revealed that most --- if not all --- discs contain substructure, includind rings and gaps \cite{ALMA_HLTau,Long2018a,Huang2018a}, spirals\cite{Benisty2015,Stolker2016a,Huang2018b}  azimuthal dust concentrations \cite{van-der-Marel2013}, and shadows cast by misaligned inner discs \cite{Marino2015,Stolker2016a}. These features have been interpreted as signatures of newborn protoplanets, but the exact origin is unknown.
Here we report the kinematic detection of a few Jupiter-mass planet located in a gas and dust gap at 130\,au in the disc surrounding the young star HD~97048. An embedded planet can explain both the disturbed Keplerian flow of the gas, detected in CO lines, and the gap detected in the dust disc at the same radius.
While gaps appear to be a common feature in protoplanetary discs\cite{Long2018a,Huang2018a}, we present a direct correspondence between a planet and a dust gap, indicating that at least some gaps are the result of planet-disc interactions.
\end{abstract}

A variety of mechanisms have been proposed to explain the formation of rings and gaps in discs, \emph{e.g.} snow-lines, non-ideal MHD effects, zonal flows, and self-induced dust-traps \cite{Takahashi2014,Gonzalez2015,Loren2015,Zhang2015,Bethune2016}. The most straightforward explanation is that the gaps are the results of forming planets interacting with the disc \cite{Dipierro15b}.
Recent ALMA surveys suggest that planets could indeed be responsible for carving out several of the observed gaps \cite{Long2018a,Huang2018a,Zhang2018a},
but until now definite evidence has remained elusive.
Despite much effort, direct imaging of planets in young discs remains difficult. Many of the claimed detections have been refuted, or require confirmation \cite{Rameau2017,
  Ligi2018,Currie2019}. The most promising detection to date is a companion imaged in the cleared inner disc around PDS~70 \cite{Keppler2018,Muller2018,Christaens2019,Haffert2019}. However, the mass estimate from photometry remains uncertain, and it is not yet clear if PDS~70~b falls within the planetary regime.

Our approach is to search for the dynamical effect of a planet on the surrounding gas disc.
Disc kinematics are dominated by Keplerian rotation. Embedded planets perturb the gas flow in their vicinity, launching spiral waves at
Lindblad resonances both inside and outside their orbit.
The disturbed velocity pattern is detectable by high spectral and spatial resolution ALMA line observations
 \cite{Perez2015b}. This technique was used to detect embedded planets in the disc surrounding HD~163296 \cite{Pinte2018b,Teague2018a}.

 Here, we used ALMA to observe the disc surrounding the young ($\approx$3\,Myr) intermediate-mass (2.4\,M$_\mathrm{\odot}$) star HD~97048 in Band 7 continuum  (885\,$\mu$m) and in the $^{13}$CO J=3-2 transition, with a spectral resolution of 220\,m\,s$^{-1}$.
 Observations were performed with 3 interferometer configurations sampling baselines from 15 to $\approx$ 8,500\,m, resulting in final images with a spatial resolution of 0.07''$\times$0.11'' (13$\times$20\,au).

 We report the detection of a localised deviation from Keplerian flow in the disc. The velocity kink is spatially associated with the gap seen in dust continuum emission (Fig.~\ref{fig:obs} and S\ref{sup_fig:SPHERE}). The most plausible and simultaneous explanation for those two independent features is the presence of an embedded body of a few Jupiter masses which carves a gap in the dust disc and locally perturbs the gas flow.

The continuum emission shows a system of two rings detected up to $\approx$ 1'' from the star. The $^{13}$CO emission extends further in radius ($\approx$ 4''), displaying the typical butterfly pattern of a disc in Keplerian rotation (Fig.~\ref{fig:obs} and \ref{fig:schema}). No significant brightness variation of the $^{13}$CO emission is detected at the location of the gap. In a given spectral channel, the emission is distributed along the corresponding isovelocity curve, \emph{i.e.} the region of the disc where the projected velocity towards the observer is equal to the channel velocity. The observed East-West asymmetry is characteristic of an optically thick emitting layer located above the midplane.
The lower, fainter, disc surface is also detected to the West of the upper disc surface.

The CO emission displays a kink in the upper isovelocity curve, highlighted by the dotted circle in Fig.~\ref{fig:obs}.
The velocity kink is detected consistently in channels between +0.7 and +1.1\,km/s from the systemic velocity. It is also seen in images reconstructed from individual observing nights, \emph{i.e.} before combining the data sets. The morphology of the emission around the velocity kink is the same with and without continuum subtraction, indicating that the kink is not the result of optical depth effects (see Supplementary Figure~\ref{sup_fig:executions}).
The sensitivity of the ALMA observations allows us to detect the continuum in each individual channel, revealing that the velocity kink is located just above the gap seen in continuum emission, at the same radius. This spatial coincidence points to a common origin for both features.
The deformation of the emission is localised to a diameter of $\approx$ 0.3''.
Notably, the emission on the opposite side the disc (and at opposite velocity) displays a smooth profile, with no kink. This excludes a large scale perturbation of the disc or an azimuthally symmetric mechanism. The perturbation is similar to the one detected in HD~163296 \cite{Pinte2018b}. In both cases, the kink is only detected over a small range in both radial extent and velocity.
A corresponding velocity kink in the lower surface of the disc is not seen as the emission
is weaker and masked by the continuum and brighter upper CO surface (Fig.~\ref{fig:obs} and \ref{fig:schema}).
Using the same procedure as in ref.\cite{Pinte2018a}, we measured the altitude of the $^{13}$CO layer to be 17$\pm$1\,au near the velocity kink (at a distance of 130\,au).
Assuming that the planet is located in the disc
midplane and exactly below the center of the velocity kink, it would be at a projected distance of 0.45$\pm$0.1" and PA =  $-55\pm$10$^\circ$ from the star.

To infer the mass of the putative planet, we performed a series of 3D global gas and multi-grain dust hydrodynamics simulations, where we embedded a planet on a circular orbit at 130\,au with a mass of 1, 2, 3 and 5\,M$_\mathrm{Jup}$ (gas disc mass of $10^{-2}$\,M$_\mathrm{\odot}$). Simulations were performed for approximately 800 orbits ($\approx$ 1\,Myr), and then post-processed to compute the thermal structure and resulting continuum emission and CO maps.

The presence of a few Jupiter mass planet produces distinct signatures in the gas and dust (Fig.~\ref{fig:phantom}). The embedded planet generates a
gap and spirals in the gas, resulting in a non-axisymmetric velocity field.
The dynamics of the dust depends on the \emph{Stokes number}, i.e. the ratio of the gas drag stopping time to the orbital time, which depends on the grain size and dust properties. When the Stokes number is close to unity --- corresponding roughly to millimetre sized grains at the gas surface densities considered here if grains are compact and spherical --- dust grains form axisymmetric rings inside and outside of the planet orbital radius \cite{Dipierro15b}.

Fig~\ref{fig:model} shows the predicted emission for the various planet masses, in the continuum and for the $^{13}$CO line. The channel maps are best reproduced with an embedded planet of 2-3\,M$_\mathrm{Jup}$, giving a velocity kink with amplitude matching the observations. For the 1\,M$_\mathrm{Jup}$ planet, the kink is too small. The most massive planet, with 5\,M$_\mathrm{Jup}$, creates a kink too large, and which remains detectable over a range of velocity that is too wide ($\pm$ 1\,km/s from the 0.96\,km/s channel where the deviation is the strongest).
Embedded planets have also been predicted to generate vertical bulk motions and turbulence which should result in detectable line broadening when the planet is massive enough ($>$ a few Jupiter masses) \cite{Dong2019}. Analysis of the moment-2 map does not reveal significant line broadening at the location of the gap and are consistent with thermal broadening and Keplerian shearing within the beam.
This also rules out the upper end of the range tested in our simulations.
HD~97048 was observed with SPHERE on the VLT, resulting in a point source detection limit of $\approx$2\,M$_\mathrm{Jup}$ \cite{Ginski2016} at the location where we detect the velocity kink. This upper limit assumes a hot-start model, and an unattenuated planet atmosphere. Our simulations show that the planet is embedded, with an optical depth of $\gtrsim$0.5 towards the observer at 1.6\,$\mu$m, \emph{i.e.} it would appear about twice as faint as an unobscured planet. This is consistent with a 2-3\,M$_\mathrm{Jup}$ planet not being detected by SPHERE. Our kinematic mass determination is also consistent with the planet mass range  0.4 and 4\,M$_\mathrm{Jup}$ estimated from the width of the scattered light gap, for a viscosity between $10^{-4}$ and $10^{-2}$ \cite{Dong2017}.

All of the planet masses explored in our models result in azimuthally symmetric gaps in continuum emission at 885\,$\mu$m, as detected by ALMA.
At this wavelength, the thermal emission is dominated by dust grains a few hundred microns in size, that decouple from the gas, and form axisymmetric rings, even if the gas flow is locally non-axisymmetric.

The width and/or depth of a gap in sub-millimetre thermal emission depends on the planet mass as well as on the Stokes number of the dust grains that contribute most at the observed wavelength \cite{Zhang2018a}.
In most cases, the Stokes number is unknown as the local gas density and dust properties are poorly constrained by observations. Continuum gap width may not provide reliable estimate of planet masses \cite{Rosotti2016}.
Conversely, when the mass of the planet is known, for instance via the kinematics of the gas as in this work,  the dust continuum observations can be used to directly measure the Stokes number, thus constraining the gas density and dust properties in the vicinity of the gap. For HD~97048, dust grains dominating the thermal emission must have a Stokes number around a few $10^{-2}$ to reproduce the observed gap profile. This excludes compact dust grains with the assumed density in our models.
 One way to reduce the Stokes number is to increase the gas density, but this causes significant accretion on the planet (see supplementary material).
Another possibility is that dust grains of a few 100 microns or millimetres in size consist of fluffy aggregates, as suggested by sub-millimetre polarisation studies \cite{Kataoka2016,
  Dent2019}. Aggregates have a larger projected area, and experience stronger gas drag than equal mass compact grains. They have smaller Stokes number, and can reproduce the observed dust continuum gap width for a 2\,M$_\mathrm{Jup}$ planet (Fig.~\ref{fig:model} and S\ref{sup_fig:porous_grains}).

The coincident location of the velocity kink and gap demonstrates that protoplanets are responsible for \emph{at least some} of the observed gaps in discs. Most of the alternative mechanisms for creating dust gaps in discs --- including snow lines, non-ideal MHD, zonal flows and self-induced dust traps --- rely on the formation of a pressure bump where the dust grains can be trapped and grow further. While those pressure bumps produce deviations from Keplerian velocity, they are axisymmetric. That is, they do not cause a localised, non-axisymmetric velocity deviation as observed.
Other mechanisms might be imagined to create a non-azimuthally symmetric velocity pattern in the disc. Gravitational instabilities or outer companion/flyby create spirals, but these are large-scale structures, and would not produce a velocity kink localised to a small region of the disc. Neither will they result in azimuthally symmetric dust gaps. The interaction of a few Jupiter masses planet with its surrounding disc is to our knowledge the only plausible explanation that can explain both a localised velocity kink and an azimuthally symmetric gap.
More systematic kinematic mass estimates may allow us to better connect the population of young embedded planets in discs with the known exoplanet population.

\renewcommand{\figurename}{\textbf{Figure}}
\renewcommand{\thefigure}{\textbf{\arabic{figure}}}

\begin{figure}
  \includegraphics[width=\linewidth]{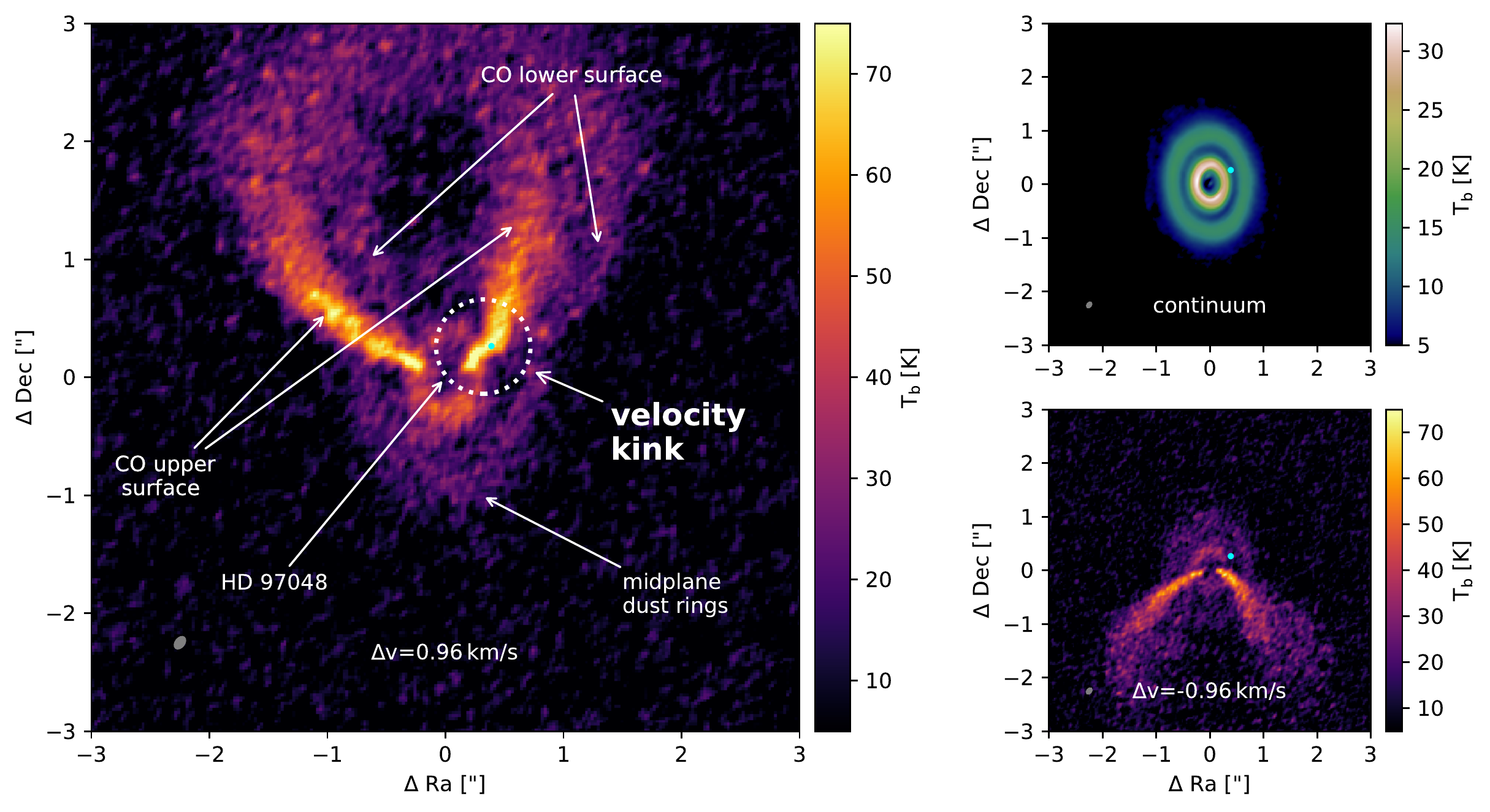}
  \caption{ALMA observations of the dust and gas disc surrounding HD~97048.
    a)  $^{13}$CO 3-2 emission at velocity +\,0.96\,km.s$^{-1}$ from the systemic velocity. The velocity kink revealing the presence of an embedded perturber is marked by a dotted circle and the cyan dot represents the location of the putative planet. The kink is located above the gap detected in continuum.
    b) 885\,$\mu$m continuum emission using all the continuum channels b) and c)  $^{13}$CO 3-2 emission at the opposite velocity -\,0.96\,km.s$^{-1}$ from the systemic velocity, where the emission displays a smooth profile. Line observations were not continuum subtracted. The ALMA beam is 0.07''$\times$0.11'' and is indicated by the grey ellipse. \label{fig:obs}}
\end{figure}

\begin{figure}
  \centering
  \includegraphics[width=0.7\linewidth,angle=0]{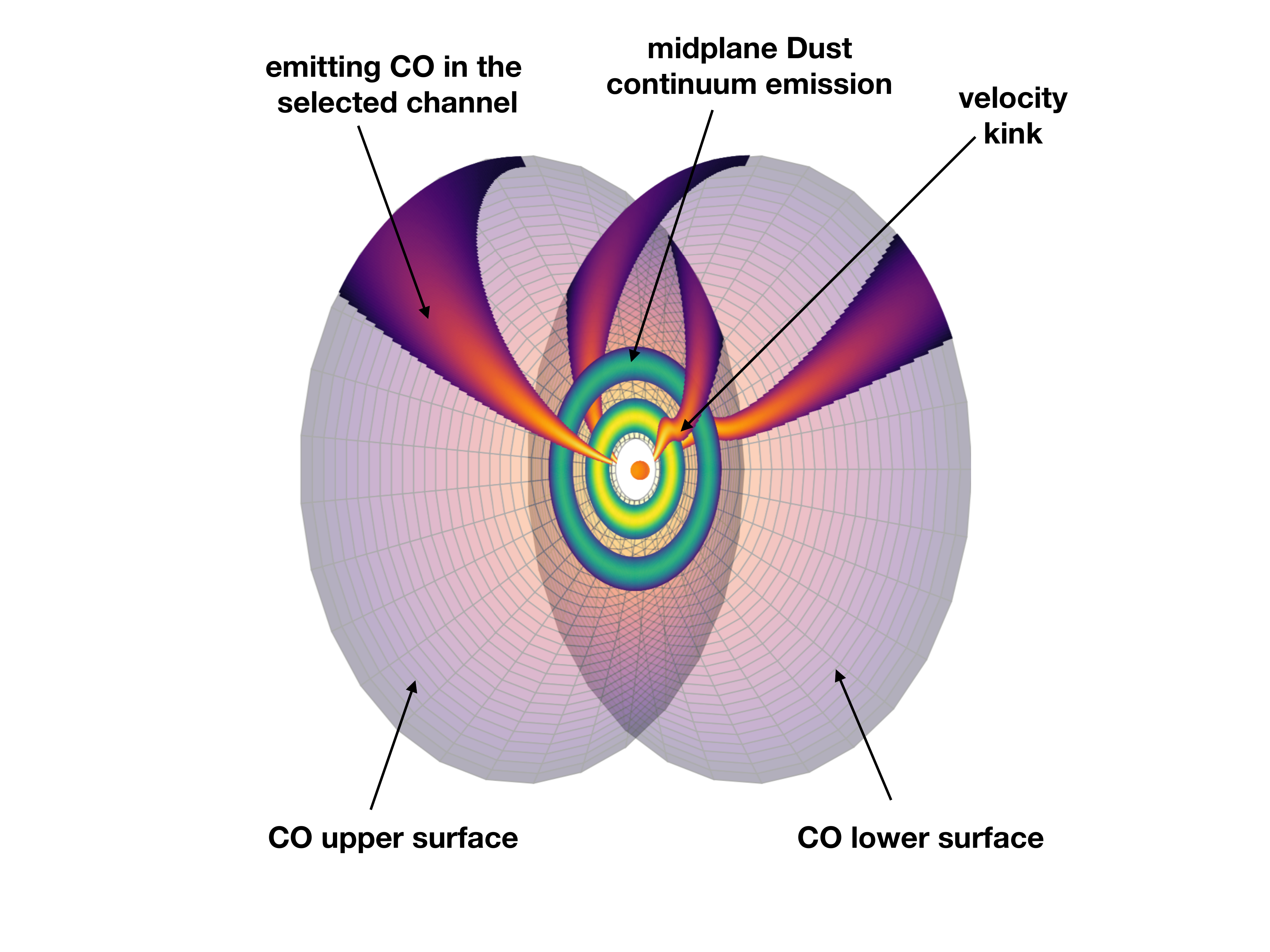}
  \caption{Schematic view of the disc as seen by ALMA in a single channel. The CO emission originates from the disc surfaces, while the continuum is mostly emitted from the disc midplane.\label{fig:schema}}
\end{figure}

\begin{figure}
  \centering
  \includegraphics[height=0.3\hsize]{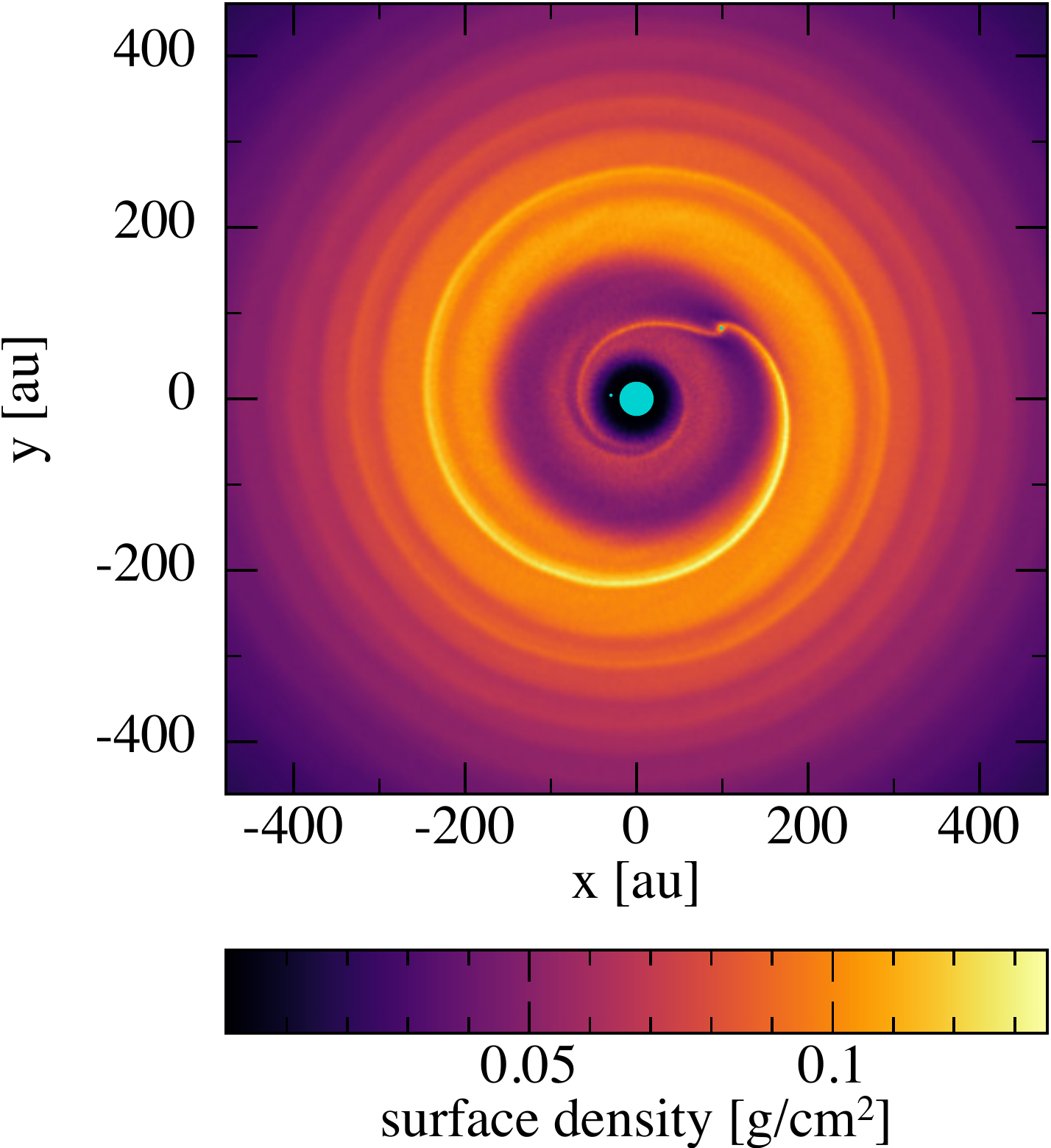}
  \includegraphics[height=0.3\hsize]{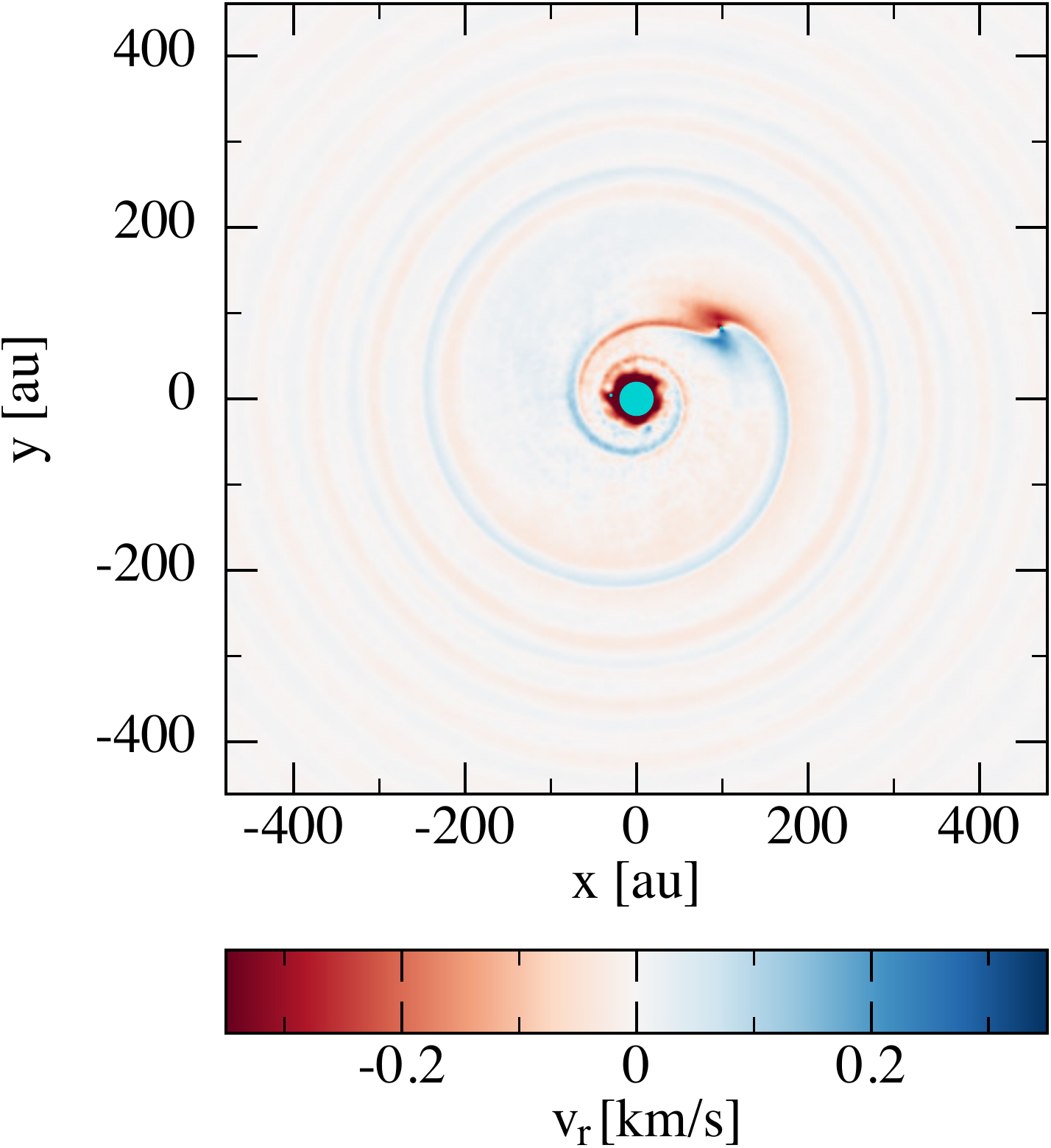}
  \includegraphics[height=0.3\hsize]{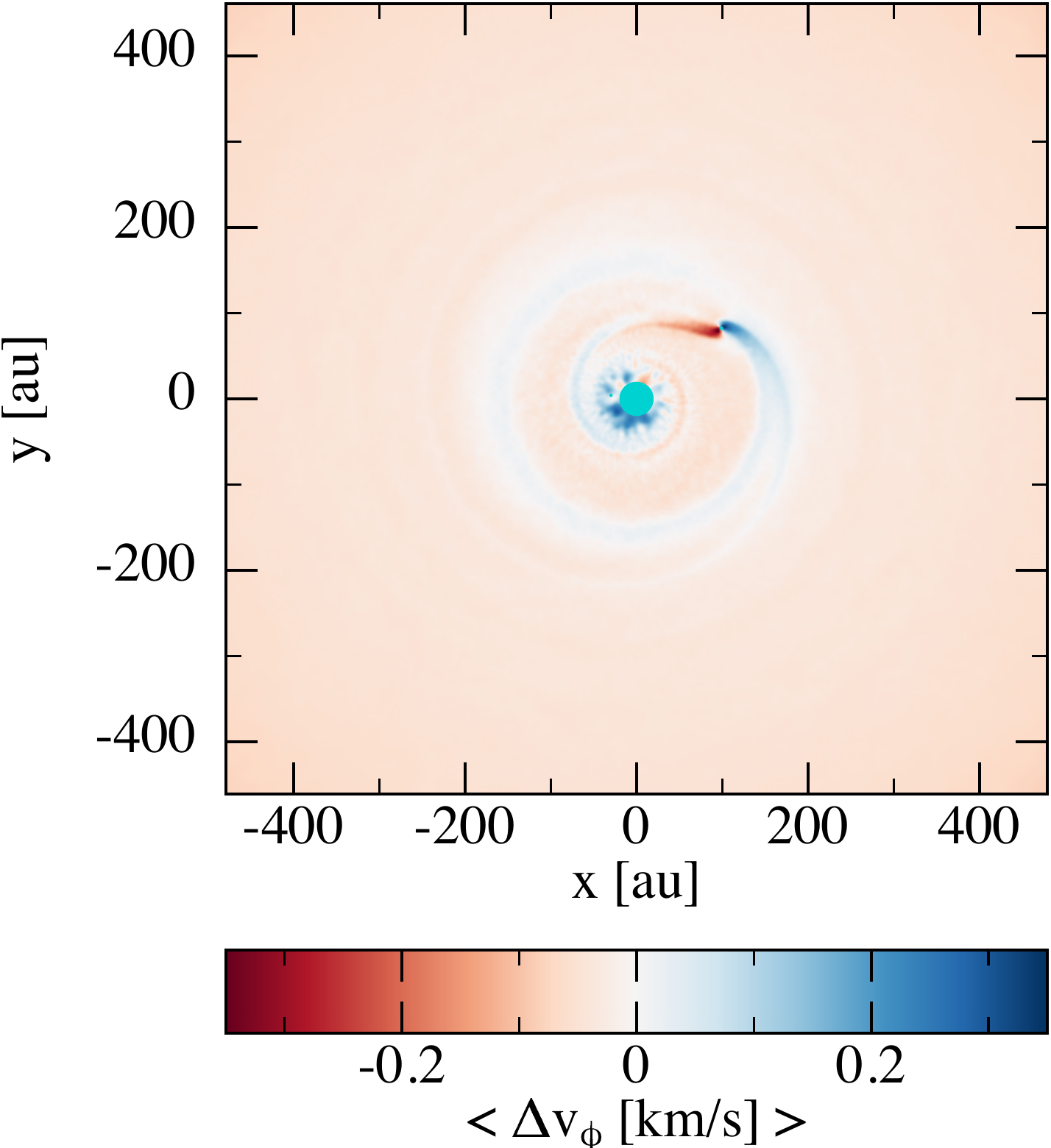}

  \vspace{2mm}
  \includegraphics[height=0.3\hsize]{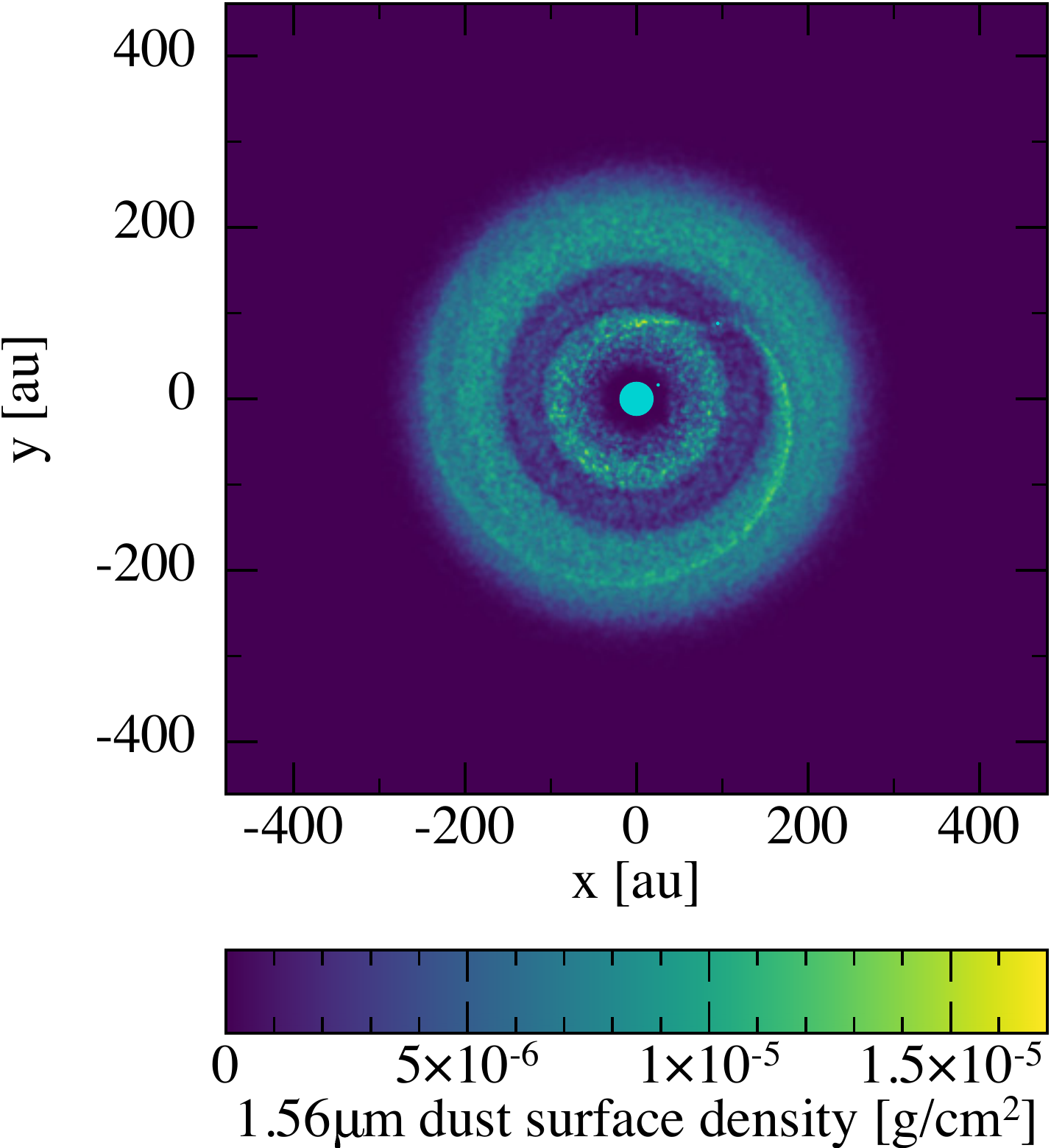}
  \includegraphics[height=0.3\hsize]{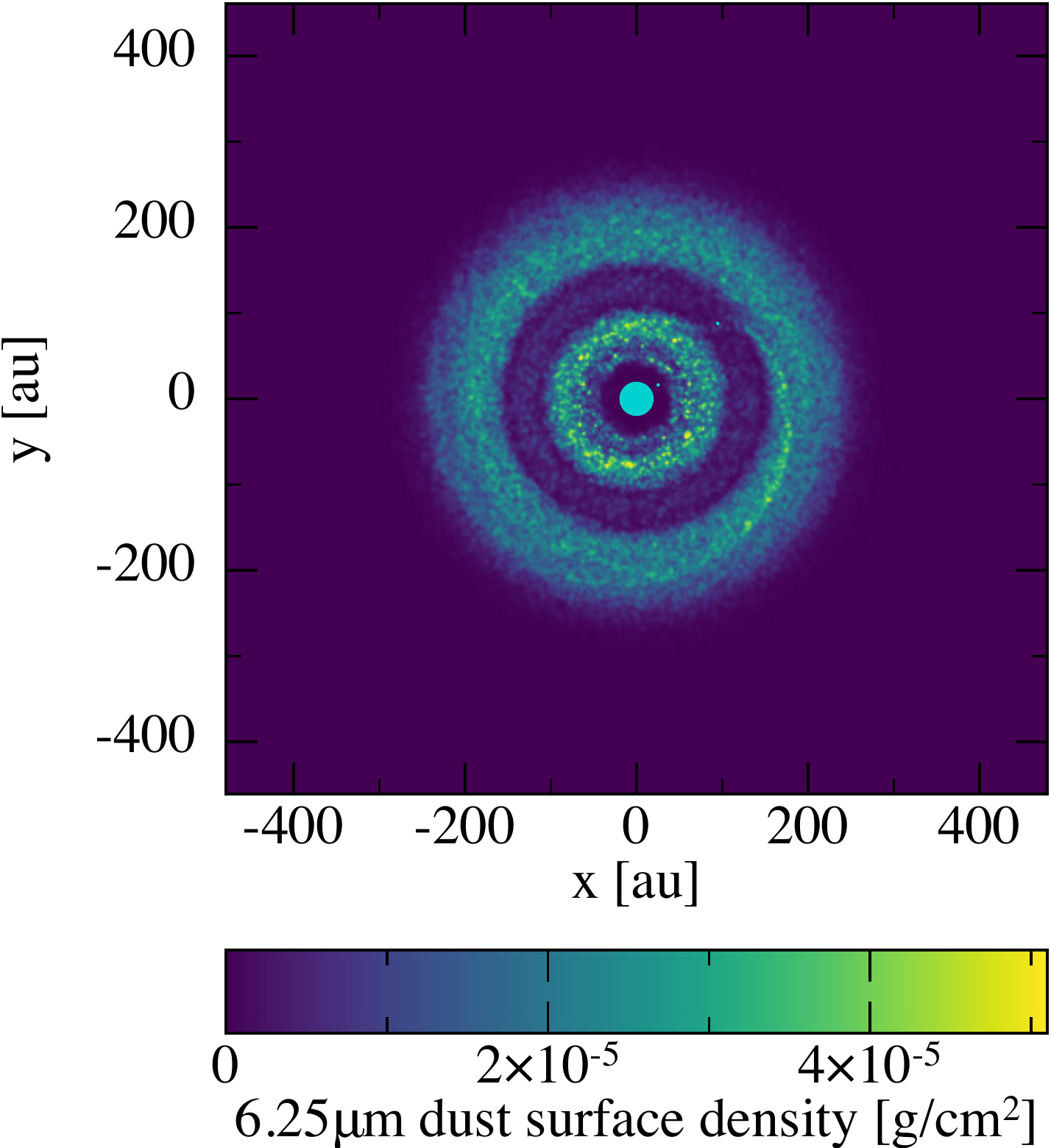}
  \includegraphics[height=0.3\hsize]{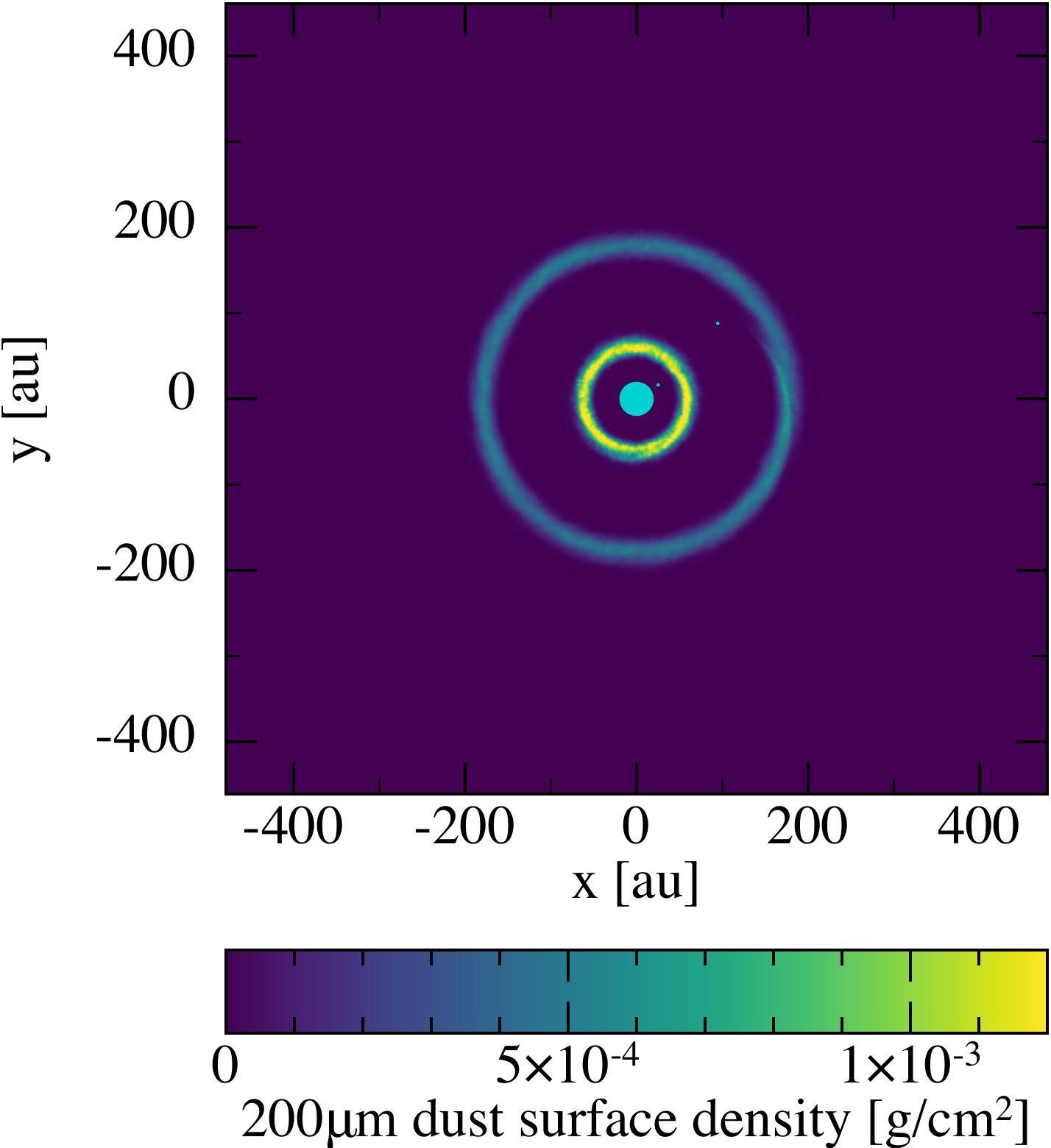}
  \caption{Hydrodynamical model of a 2\,M$_\mathrm{Jup}$ planet interacting with the disc of HD~97048. \emph{Top row:} Gas surface density (left), gas radial velocity (middle) and gas azimuthal velocity offset compared to Keplerian velocity (right). \emph{Bottom row:} Dust surface density for 1.5\,$\mu$m (left), 6.25\,$\mu$m (middle) and 200\,$\mu$m (right) compact dust grains. They have a Stokes number of $\approx 10^{-2}$, $5 \times 10^{-2}$, and $1$, respectively. Fluffy aggregates and/or porous grains will have a smaller Stokes number for the same grain mass. Sink particles are marked by cyan dots with size corresponding to their accretion radii. \label{fig:phantom}}
\end{figure}

\begin{figure}
  \includegraphics[width=\linewidth]{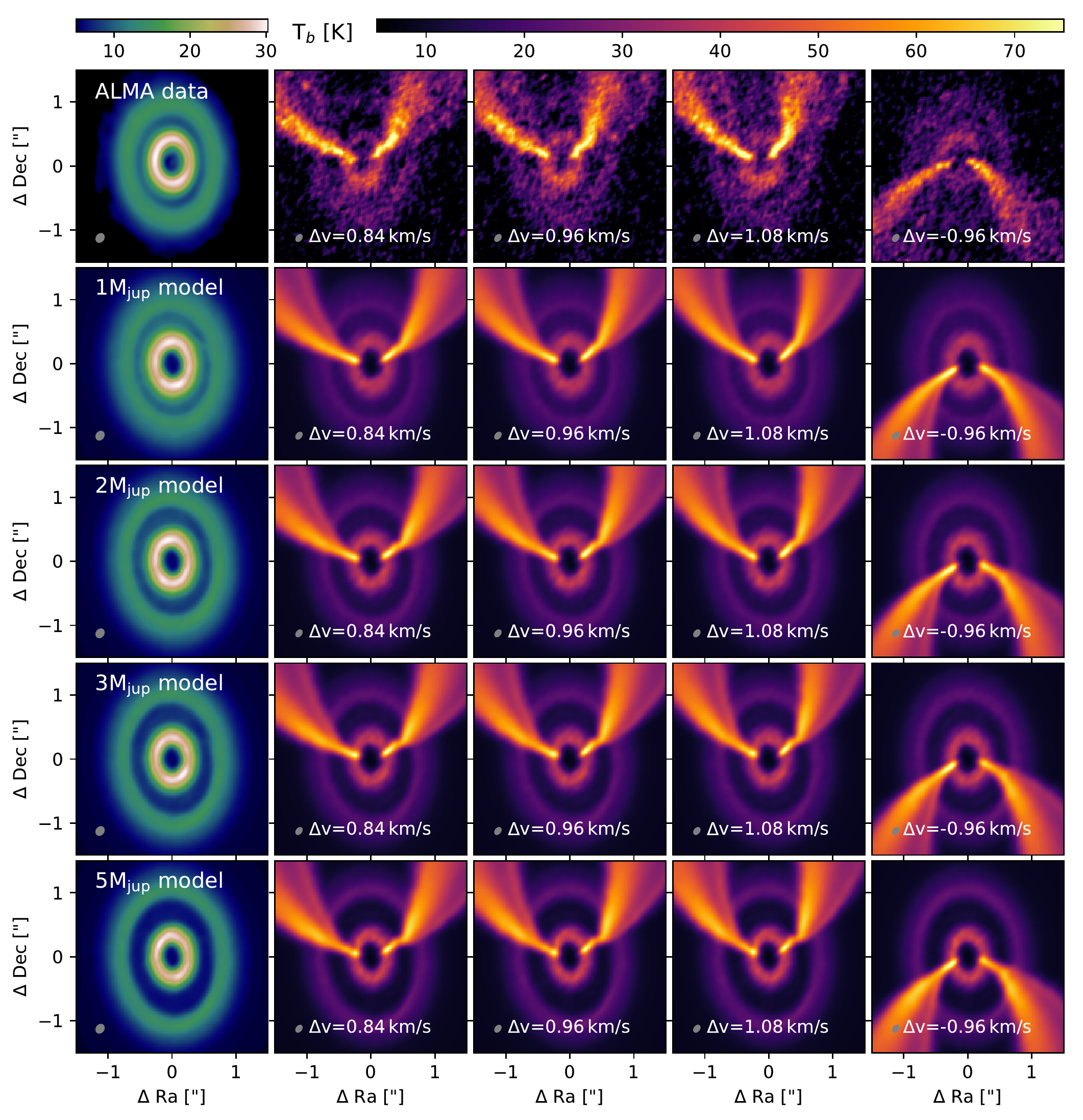}
  \caption{Comparison of ALMA observations (top row)
    with hydrodynamic simulations of discs with different embedded planet masses, post-processed with radiative transfer. The left column shows the continuum data, while the following panels show the line data. The 2\,M$_\mathrm{Jup}$ case corresponds to the model displayed in Fig.~\ref{fig:phantom}. Porous grains and/or aggregates are required to match the continuum gap width. The surface to mass ratio of the dust grains has been increased by 50 compared to compact spheres.\label{fig:model}}
\end{figure}

%% Put the bibliography here, most people will use BiBTeX in
%% which case the environment below should be replaced with
%% the \bibliography{} command.
\newcommand{\aap}{Astron. Astrophys.}
\newcommand{\apjl}{Astrophys. J. Letters}
\newcommand{\apj}{Astrophys. J.}
\newcommand{\aj}{Astron. J}
\newcommand{\mnras}{Mon. Not. R. Astron. Soc.}
\newcommand{\pasa}{Publ. Astron. Soc. Australia}
\bibliography{biblio}

%% Here is the endmatter stuff: Supplementary Info, etc.
%% Use \item's to separate, default label is "Acknowledgements"

\begin{addendum}
 \item[Competing Interests.] The authors declare that they have no
competing financial interests.
 \item[Correspondence.] Correspondence and requests for materials
   should be addressed to C.P.\break (email:  christophe.pinte@monash.edu).
 \item[Author contributions.]
   C.P. analysed the data, carried out the modelling and wrote the manuscript. G. vdP wrote the observing proposal and reduce the data. D.P. provided advice on running the smoothed particle hydrodynamics simulations and made some of the figures. All co-authors provided input on the manuscript.
    \item[Acknowledgements.] C.P., D.J.P. and V.C. acknowledge funding from the Australian Research Council via\break
FT170100040 and DP180104235. F.M., GvdP and C.P. acknowledge funding
from ANR of France (ANR-16-CE31-0013).
This work was performed on the OzSTAR national facility at Swinburne University
of Technology. OzSTAR is funded by Swinburne and the Australian Government's
Education Investment Fund.
  This paper makes use of the following ALMA data:
 ADS/JAO.ALMA\#2016.1.00826.S. ALMA is a
 partnership of ESO (representing
its member states), NSF (USA) and NINS (Japan), together with NRC
(Canada) and NSC and ASIAA (Taiwan) and KASI (Republic of Korea), in
cooperation with the Republic of Chile. The Joint ALMA Observatory is
operated by ESO, AUI/NRAO and NAOJ. The National Radio Astronomy Observatory is
a facility of the National Science Foundation operated under
cooperative agreement by Associated Universities, Inc.
 \end{addendum}

\begin{table*}
\caption{Details of the observations.\label{table_obs}} \smallskip
\begin{minipage}[t]{\textwidth}
\centering
%\noindent\begin{tabularx}{\columnwidth}{@{\extracolsep{\stretch{1}}}*{8}{l}@{}}
\begin{tabular}{lccccccc}
\hline\hline
UTC Date & Time on source & N$_{\mathrm{ant}}$ & Baselines & pwv  &  \multicolumn{3}{c}{Calibrators:} \\ % \cline{5-7}
 & [minutes] &  & [m] & [mm] &  Bandpass & Flux & Phase  \\ % gain == amplitude?
\hline
2016 Nov 18 & 26.3 & 45 & 15 to 919  & 0.60 & J0538-4405 &  J1107-4449 &  	J1058-8003\\%
2017 Nov 24 & 42.7 & 49 & 92 to 8548 & 0.56 & J0522-3627 &  J0904-5735 &  	J1058-8003\\%
2017 Nov 30 & 42.7 & 47 & 79 to 8283 & 0.56 & J0635-7516 &  J0904-5735 &  	J1058-8003\\%
\hline
\end{tabular}
\end{minipage}
\end{table*}

\newpage
\begin{methods}
%Put methods in here.  If you are going to subsection it, use
%\verb|\subsection| commands.  Methods section should be less than
%800 words and if it is less than 200 words, it can be incorporated
%into the main text.

\subsection{HD~97048.}

HD~97048 is located in the Chameleon~I cloud, at a distance of 185\,pc \cite{Gaia-Collaboration2018}.  It has a spectral type Be9.5/AO, an effective temperature of 10,000\,K, and a luminosity of 40 solar luminosities. The star is surrounded by a large disc \cite{Lagage06,van-der-Plas2017}, extending up to at least 850\,au in $^{12}$CO J=2-1 emission which is seen at an inclination $\approx 40^\circ$.
Previous ALMA observations of the dust sub-disc (probing dust grains a few hundred microns in size) revealed a central cavity and two bright rings separated by a gap centered at $\approx$130\,au  from the star \cite{van-der-Plas2017}. The two bright rings are also detected in scattered light with VLT/SPHERE \cite{Ginski2016} (Supplementary figure~\ref{sup_fig:SPHERE}), where the observations probes the distribution of sub-micron sized dust grains. Such small grains experience a high gas drag and tend to follow closely the gas spatial distribution, suggesting that the gap is also present in the gas disc. Two extra rings  extending up to 2.2'' from the star are also detected in scattered light. The disc is detected in PAH emission up to $\approx$ 650\,au, revealing a flaring surface \cite{Lagage06}. Scattered light images confirmed that the disc surface shows a significant flaring  \cite{Ginski2016}.

\subsection{Observations and data reduction.}

We observed HD~97048 with ALMA in band~7 in the C40-4 (1 execution) and C40-7 (2 executions) configurations reaching a total time on source of 112 minutes (ALMA program \#2016.1.00826.S, PI: G. van der Plas). The details of the observations can be found in Table \ref{table_obs}. One of the spectral windows for each observation was centered at the $^{13}$CO J=3-2 rest frequency with an individual channel width of 122\,kHz, resulting in a 244\,kHz spectral resolution (220\,m s$^{-1}$) after Hanning smoothing. The three other spectral windows were used for continuum with a bandwidth of 1.875\,GHz each.

We performed one round of phase self calibration on the continuum data set observed November 24th, and two rounds on the other two data sets, and applied the self calibration solutions to the line data. We imaged the visibilities at a 120\,m\,s$^{-1}$ velocity spacing using Briggs weighing,  resulting in a beam size of 0.11'' $\times$ 0.07'' arcsec at a PA of -38$^\circ$. Typical RMS values obtained are 3.6\,mJy/beam per channel for the combined data.
Channel maps are presented in Supplementary Figure~\ref{sup_fig:channel_maps}.

The kink in CO emission discussed in this manuscript is present in all 3 individual executions and in both the continuum subtracted and non-subtracted images (Supplementary Figure~\ref{sup_fig:executions}).

\subsection{3D modelling procedure.}

We performed a series of 3D global simulations using the {\sc phantom} Smoothed
Particle Hydrodynamics (SPH) code \cite{Price2018a}.
We performed a series of multi-grain gas+dust simulations using the algorithm described in \cite{Hutchison2018,Ballabio2018}, using 2 million SPH particles and following the dust fraction of particles of sizes ranging from 1.5625 to 1600\,$\mu$m
 (11 bins total, each bin doubling the grain size).
Each dust species experiences a different gas drag depending on the grain size, resulting in differential vertical setting and radial migration.
All 11 populations of dust grains were evolved simultaneously with the gas. We thus self-consistently took account of the cumulative backreaction on the gas.

We assume a central mass of 2.4\,M$_\odot$ and a distance of 185\,pc \cite{Gaia-Collaboration2018}.
We set the initial disc inner and outer radii to 40\,au and 700\,au, respectively.
We set the gas mass to $10^{-2}$\,M$_\odot$, and use an exponentially
tapered power-law surface density profile with a critical radius of 500\,au,
power-law index of $- 0.5$. The disc aspect
ratio was set to 0.06 at 40\,au (consistent with the scattered light images \cite{Ginski2016}), with a vertically isothermal equation of state, and sound speed power-law index of $-0.25$.
We set the artificial viscosity in the code to obtain an average
Shakura-Sunyaev\cite{ShakuraSunyaev1973} viscosity of $10^{-3}$ \cite{Lodato2010}.

We embedded two planets in the disc orbiting at 30\,au, with a mass of 1\,M$_{\rm Jup}$, and at 130\,au with a mass of
either 1, 2, 3, or 5\,M$_{\rm Jup}$. The inner planet is
used to carve a central cavity in the disc, as seen in the ALMA observations, but we did not vary the mass of the inner planet in this work. The presence of an inner planet is also suggested by the torqued intra-cavity HCO$^+$ velocity field \cite{van-der-Plas2017}.
We used sink particles \cite{Bate95} to
represent the star and planets. We set the accretion radius of the planets to 0.125
times the Hill radius, with an
accretion radius of 10\,au for the central star. The model surface
density is plotted in the top left panel of figure~\ref{fig:phantom} for the 2\,M$_\mathrm{Jup}$
planet, along with the radial velocity (top centre) and predicted deviation from Keplerian flow (top right panel) and dust column density for 1.56, 6.25 and 200$\mu$m grains (bottom row). We evolved the models for 800 orbits of
the outer planet ($\approx 1$ million years). The flow pattern around the planet establishes itself over a much shorter timescale, but the gap carving is set by the viscous timescale and establishment of dust gaps at lower Stokes numbers requires a large number of orbits.

The planets accrete a moderate amount of gas from the disc. The final masses after 800 orbits are 1.38, 2.52, 3.63 and 5.77\,M$_\mathrm{Jup}$ for the 1, 2, 3, 5\,M$_\mathrm{Jup}$ planets respectively. Migration is negligible with all planets migrating by less than 1\,au by the end of the calculations.

Additional simulations were also performed with a disc gas mass of $10^{-1}$\,M$_\odot$. As they result in significant accretion on the outer planet, we explore a range of planet masses. Planets with initial masses of 0.1, 0.15, 0.2, 0.25, and 0.5\,M$_\mathrm{Jup}$ reach a mass of 0.11, 0.18, 0.30, 2.0, 3.6, and 5.9\,M$_\mathrm{Jup}$, respectively, after 800 orbits. Migration remains limited to less than 3\,au for all planet masses, except for the most massive planet which migrated by 9\,au.
We can produce a velocity kink matching the observations using a planet with a final mass of 2\,M$_\mathrm{Jup}$, giving us a similar planet mass estimate as with the lower disc gas mass models.
However, in the high gas mass models the accretion rate on the planet increases with time, leading to a runaway accretion process and a high final planet mass.

To compute the disc thermal structure, continuum images and synthetic line maps, we used
 the {\sc mcfost} Monte Carlo radiative
transfer code \cite{Pinte06,Pinte09}, assuming
$T_\mathrm{gas} = T_\mathrm{dust}$, and local thermodynamic equilibrium as we are looking at low-$J$ CO lines. The central star was represented by a sphere
of radius 2.25\,R$_\odot$, radiating isotropically with a Kurucz
spectrum at 10,000\,K \cite{Woitke2018a}.
To avoid interpolating the density structure between the SPH and radiative transfer code,
we used a Voronoi tesselation where each {\sc mcfost} cell corresponds
to a SPH particle.
We set the $^{13}$CO abundance to 7\,10$^{-7}$ and follow the prescription described in
Appendix B of \cite{Pinte2018a} to account for freeze-out where T $< 20$\,K, photo-dissociation and photo-desorption in locations where the UV radiation is high. We adopted a turbulent velocity of 50 m\,s$^{-1}$.

We adopted a fixed dust mixture composed of 60\,\%
silicate, 15\,\%
amorphous carbon \cite{Woitke2016}. Each grain size is represented by a
distribution of hollow spheres with a maximum void fraction of $0.8$.
  We use a grain population with 100 logarithmic bins in size ranging from 0.03 to 3000$\mu$m.
At each point in the model, the density of a given grain size is obtained by interpolating between the SPH dust grain sizes, assuming grains smaller than half the smallest SPH grain size, ie \,0.78$\mu$m, follow the gas distribution, and grains larger than 1.6\,mm follow the distribution of the 1.6\,mm grains.
The total grain size distribution is normalised by integrating over all grain sizes, assuming  a power-law
$\mathrm{d}n(a) \propto a^{-3.5}\mathrm{d}a$, and over all cells, where we set the total dust mass to 1/100 of the total SPH gas mass. We computed the dust optical properties using the Mie theory.

CO maps were generated at a spectral resolution of 50\,m/s, binned at the observed resolution and Hanning smoothed to match the observed spectral resolution. Continuum and CO maps were then convolved with a Gaussian beam matching the ALMA CLEAN beam.

The spatial distribution of the dust grains is set by their Stokes number. In {\sc phantom}, we assume compact spheres, but grains with different shape and mass will follow the same spatial distribution if they have the same Stokes number.
  To study the impact of the Stokes number on the thermal emission maps, we can mimic fluffyness by shifting the SPH grain sizes before interpolating the grain size distribution in {\sc mcfost}.
Thermal emission at 885\,$\mu$m is dominated by dust grains a few hundred microns in size. For our model with a 10$^{-2}$\,M$_\odot$ gas mass, compact grains of this size would have a Stokes number close to 1. The corresponding gas width appears larger than in the observations.
Supplementary Figure \ref{sup_fig:porous_grains} shows that if the emitting grains have a Stokes number around $10^{-2}$ instead of $1$, for instance if they are fluffy aggregates, the sub-millimetre gap and ring widths are in much better agreement with the observations for the same mass planet.

\subsection{Radiative-equilibrium hydrodynamics calculations.}

  The $^{13}$CO emission originates from between 1 and 2 hydrostatic scale heights\cite{Pinte2018a}, at an altitude where the temperature is higher than in the midplane. To test the validity of the vertically isothermal structure used in the SPH calculations, we also performed a set of {\sc phantom} simulations where the temperature is regularly updated by {\sc mcfost}. The two codes have been interfaced to run simultaneously. Thanks to the fast mapping between the distribution of SPH particles and radiative transfer Voronoi mesh,
we can perform frequent radiative transfer calculations within the SPH simulation.
At a specified time interval, {\sc phantom} passes the local density, grain distribution, and sink particles properties to {\sc mcfost} which returns the 3D disc temperature (assuming the gas temperature is equal to the dust temperature). Between two {\sc mcfost} calls, {\sc phantom} evolves the disc with the temperature of each particle held constant.
 This main advantage of this method is that we include the full frequency dependence, as well as light scattering, which are critical for accurate temperature calculations in protoplanetary discs.
We assume radiative equilibrium at each call of {\sc mcfost}, which is only a valid approximation if the radiative timescale is much smaller than the dynamical timescale. Due to the limited optical depths of the models of HD~97048, these conditions are satisfied here. The temperature structure is updated every $1/10^{th}$ of the outer planet orbit. Simulations were also performed with calls to {\sc mcfost} once per orbit, producing almost indistinguishable results.

Supplementary figures \ref{sup_fig:vphi}, \ref{sup_fig:vr}  and \ref{sup_fig:vz} compares the velocity fields for the 2\,M$_\mathrm{Jup}$ model between the vertically isothermal and vertically stratified (\emph{ie} with regular {\sc mcfost} temperature updates) cases. The isothermal simulation was designed to have
a similar midplane temperature as the {\sc mcfost+phantom} model, \emph{i.e.} with $h/r = 0.06$ at $r=40$\,au. This is the same simulation as presented in the rest of the paper.
As expected, differences increase with altitude, with larger deviations in the vertically stratified case.
In this model, they remain limited to $\approx$0.1 and 0.05\,km/s for the azimuthal and radial velocities, respectively.

In the case of the $^{13}$CO emission of HD~97048, the vertically isothermal structure appears as a reasonable approximation, as long as the $h/r$ is chosen sensibly. Hence vertical temperature stratification does not significantly affect our planet mass estimate.

\subsection{Impact of observational noise and $uv$ plane sampling.}

As we do not aim to perform a detailed fitting of the data, all models so far were presented with a simple Gaussian convolution to compare with observations, \emph{i.e.} showing noise-free and with fully-sampled $uv$-plane synthetic maps.

  To assess if observational artefacts could affect the images and in particular the detection of the kink, we also post-processed the 2\,M$_\mathrm{Jup}$ model through a modified version of the {\sc CASA} ALMA simulator. Synthetic visibilities were computed at the same ($u$,$v$) coordinates as the data. A precipitable water vapor of 0.6\,mm was used to set the thermal noise.
The resulting synthetic visibilities were CLEANed using the same
parameters as  observed visibilities.

A comparison of the Gaussian-convolved and CLEANed synthetic images is showed in supplementary figure~\ref{sup_fig:test_conv}.
The shape of the velocity kink is not affected by observational effects, indicating that a simple convolution is a good approximation for qualitatively comparing models to data.

\end{methods}

\section*{Data availability}
Raw data is publicly available via the ALMA archive under project id 2016.1.00826.S. Final reduced and calibrated data cubes are available with the DOI 10.6084/m9.figshare.8266988.

\section*{Code availability}
{\sc Phantom} is publicly available at \url{https://bitbucket.org/danielprice/phantom}. {\sc mcfost} is currently available under request and will be made open-source soon.
 Figures were generated with {\sc splash}\cite{Price2007} (\url{http://users.monash.edu.au/~dprice/splash/}) and {\sc pymcfost} (\url{https://github.com/cpinte/pymcfost}), which are both open-source.

\newpage
\setcounter{page}{1}
\setcounter{figure}{0}

\renewcommand{\figurename}{\textbf{Supplementary Figure}}
\renewcommand{\thefigure}{\textbf{\arabic{figure}}}

\begin{figure}
  \centering
  \includegraphics[width=0.7\linewidth,angle=0]{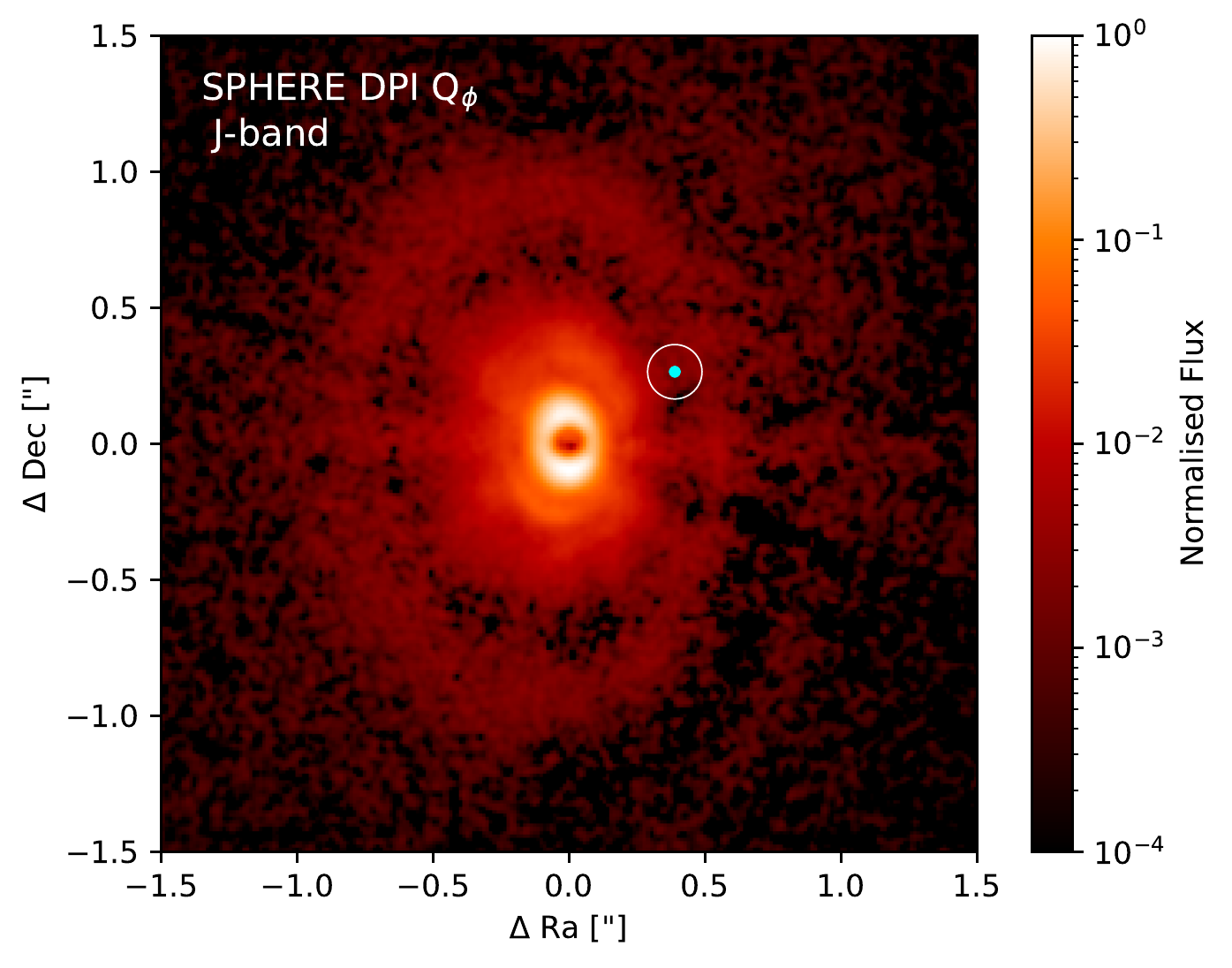}
  \caption{VLT/SPHERE dual-polarisation image Q$_\phi$ image in J band (data from \cite{Ginski2016}), with the cyan dot marking the location of the putative planet. The white circle represents an uncertainty of 0.1'' on the planet location. The data was convolved with a Gaussian of FWHM 2 pixels (24.5\,mas) to reduce noise.\label{sup_fig:SPHERE}}
\end{figure}

\begin{figure}
  \centering
  \includegraphics[width=\linewidth,angle=0]{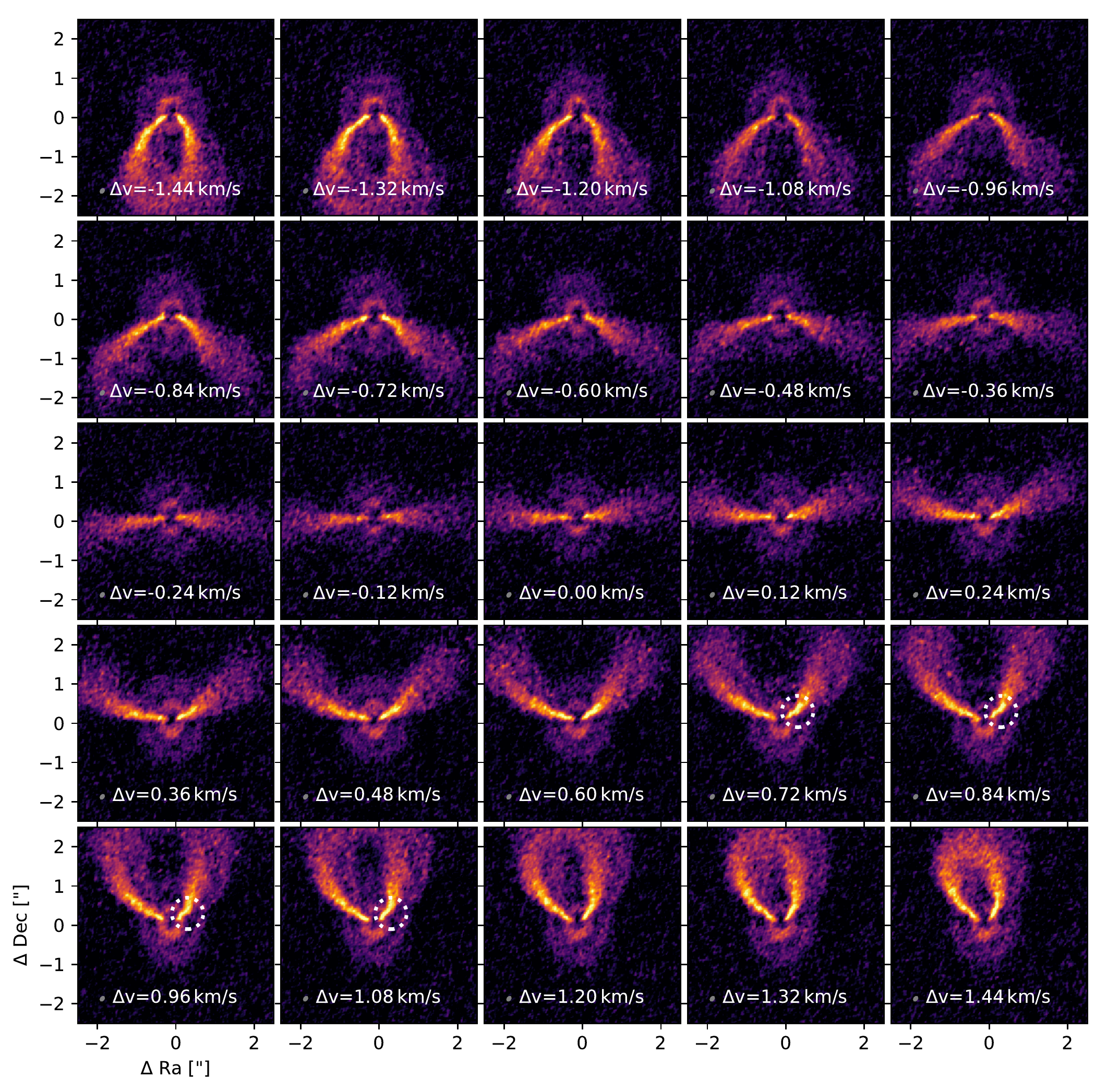}
  \caption{$^{13}$CO channel maps of HD~97048. The velocity kink is only visible in a few channels between 0.7 and 1.1\,km/s from the systemic velocity (marked by a dashed circle).\label{sup_fig:channel_maps}}
\end{figure}

\begin{figure}
  \centering
  \includegraphics[width=0.7\linewidth,angle=0]{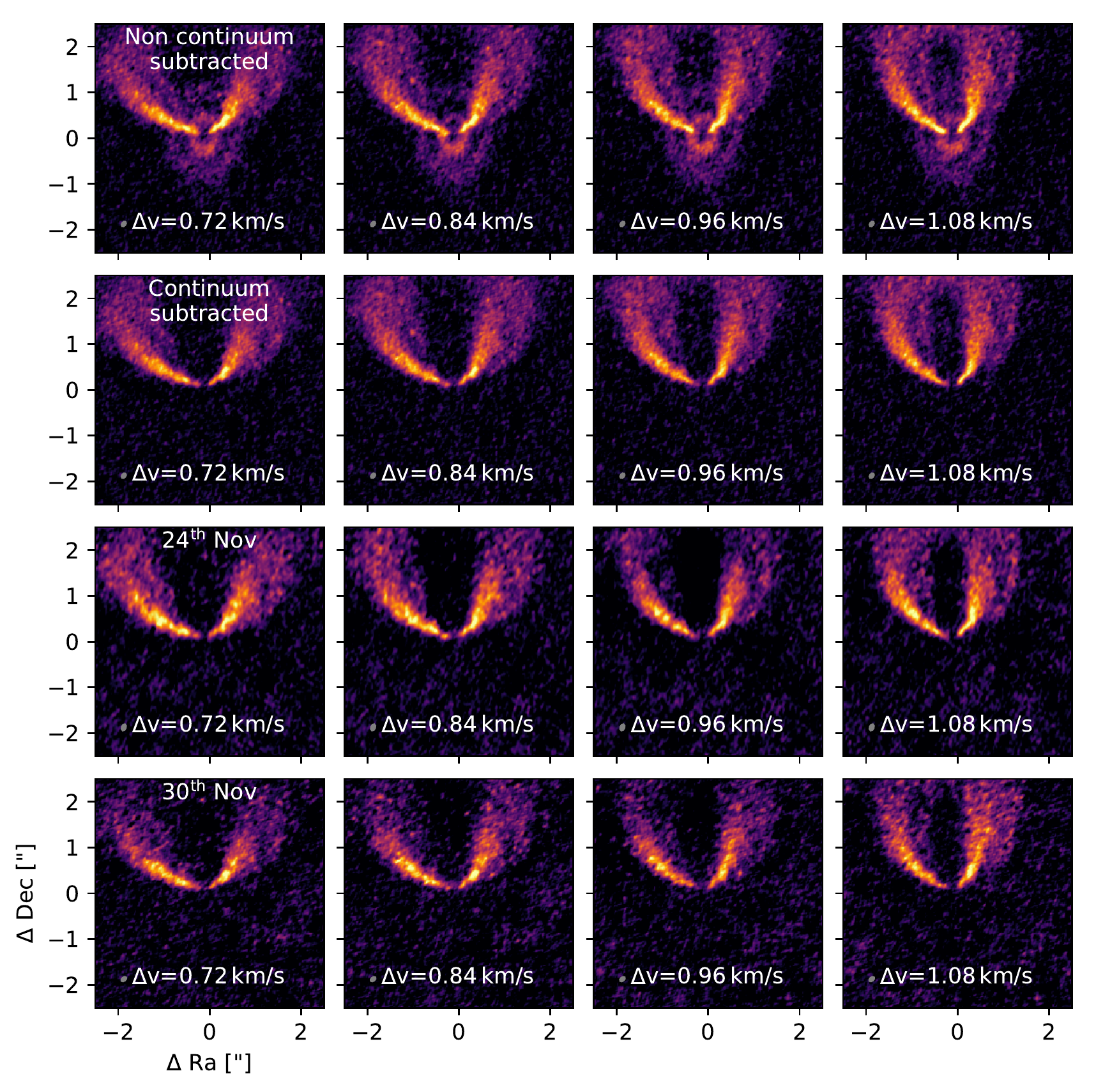}
  \caption{Imaging with and without continuum subtraction, and of the two individual datasets with extended ALMA configuration. The velocity kink is consistently detected in both data sets, and the continuum subtraction does not affect the detection.\label{sup_fig:executions}}
\end{figure}

\begin{figure}
  \centering
  \includegraphics[width=0.7\linewidth,angle=0]{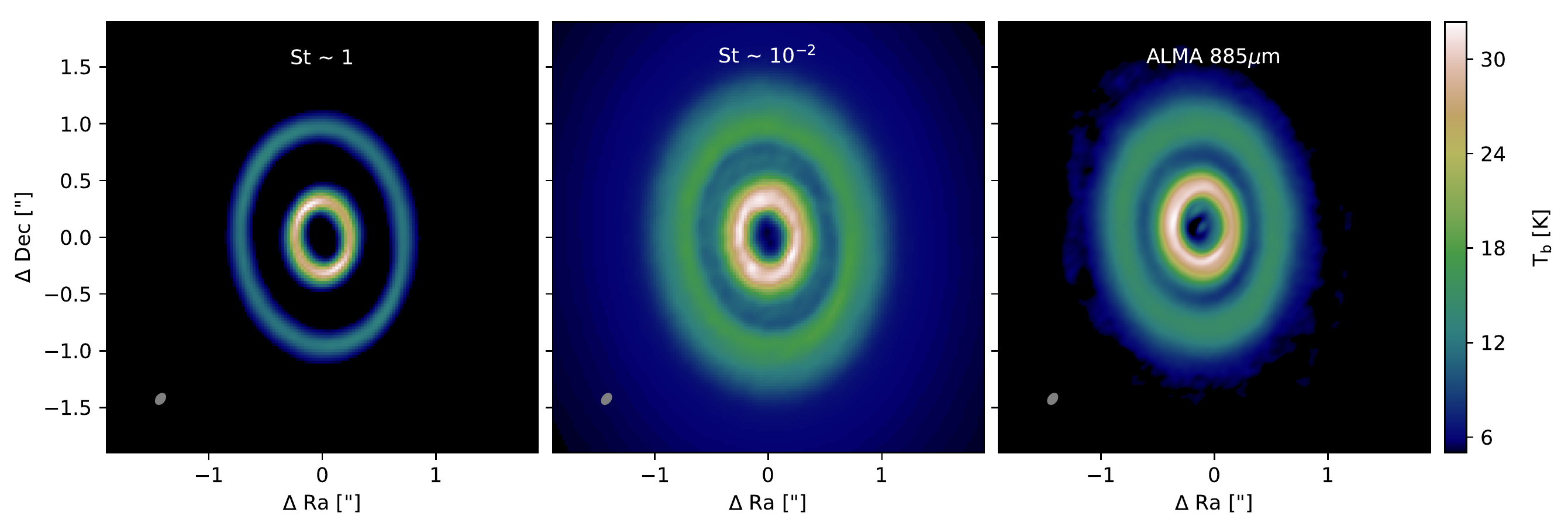}
  \caption{Continuum emission at 885\,$\mu$m as a function of the Stokes number of the 200\,$\mu$m grains. \emph{Left:} The 200\,$\mu$m grains have a Stokes number close to 1 in the rings. This is the reference model presented in this paper. \emph{Center:} The 200\,$\mu$m grains have a Stokes number close to $10^{-2}$. \emph{Right:} ALMA continuum observations. \label{sup_fig:porous_grains}}
\end{figure}

\begin{figure}
  \includegraphics[width=\linewidth]{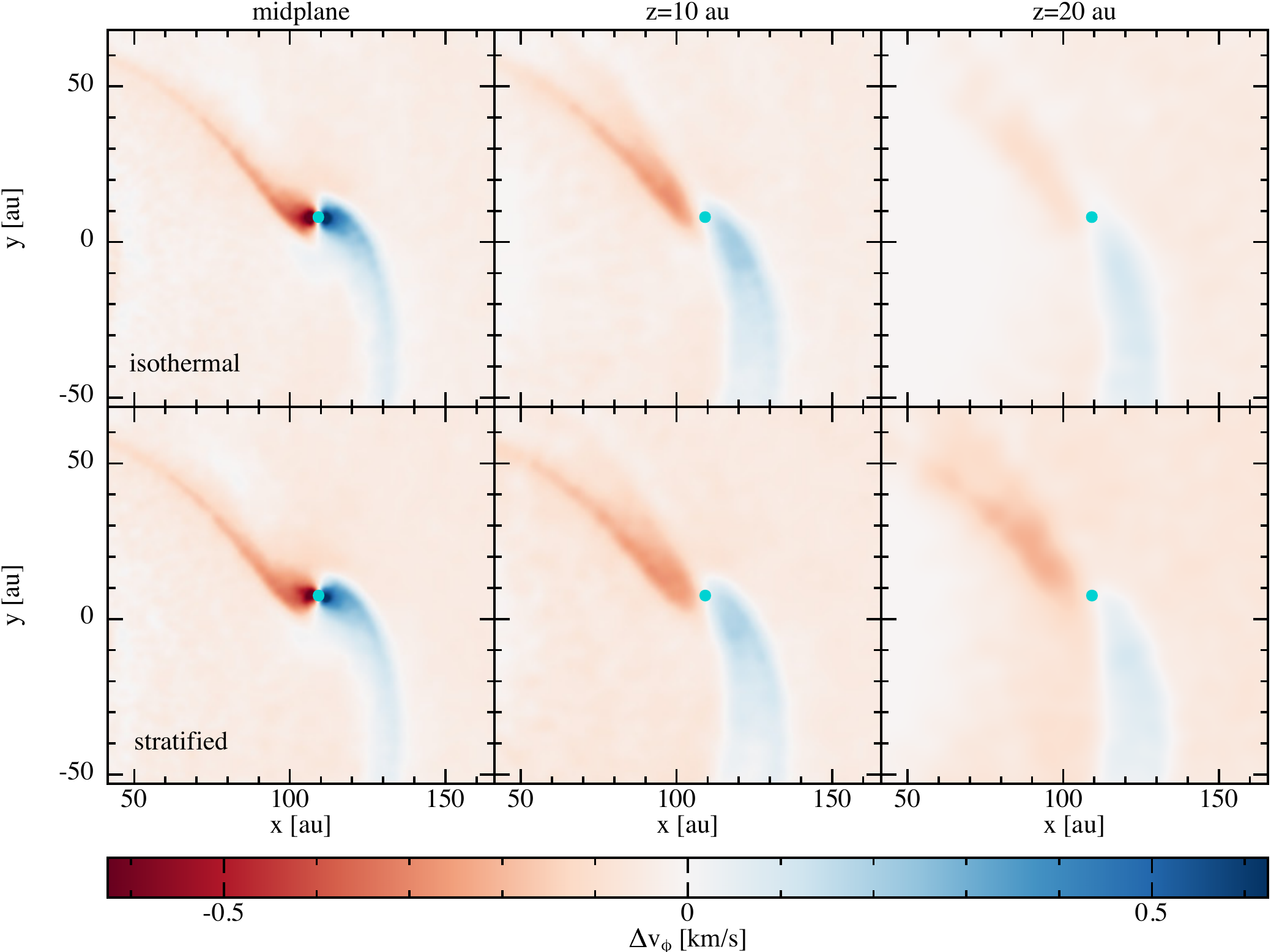}
\caption{Azimuthal velocity perturbation induced by a 2M$_\mathrm{Jup}$ planet in a vertically isothermal calculation (top) compared to a calculation with stratified vertical temperature (bottom). We show slices taken in the midplane (left) and at heights of 10 and 20 au above the midplane (middle and right, respectively). \label{sup_fig:vphi}}
\end{figure}

\begin{figure}
  \includegraphics[width=\linewidth]{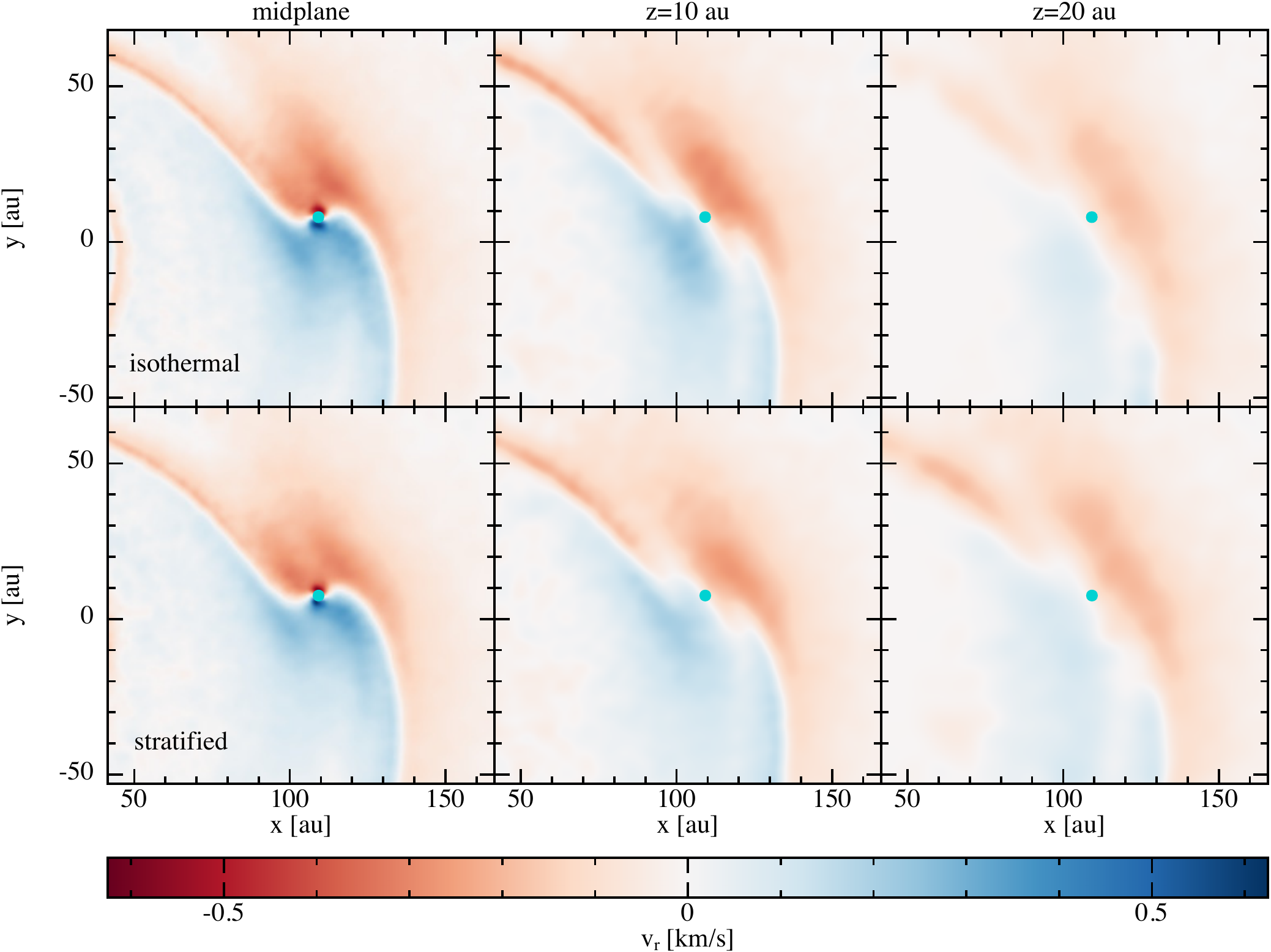}
\caption{Radial velocity perturbation induced by a 2M$_\mathrm{Jup}$ planet in a vertically isothermal calculation (top) compared to a calculation with stratified vertical temperature (bottom). We show slices taken in the midplane (left) and at heights of 10 and 20 au above the midplane (middle and right, respectively). \label{sup_fig:vr}}
\end{figure}

\begin{figure}
  \includegraphics[width=\linewidth]{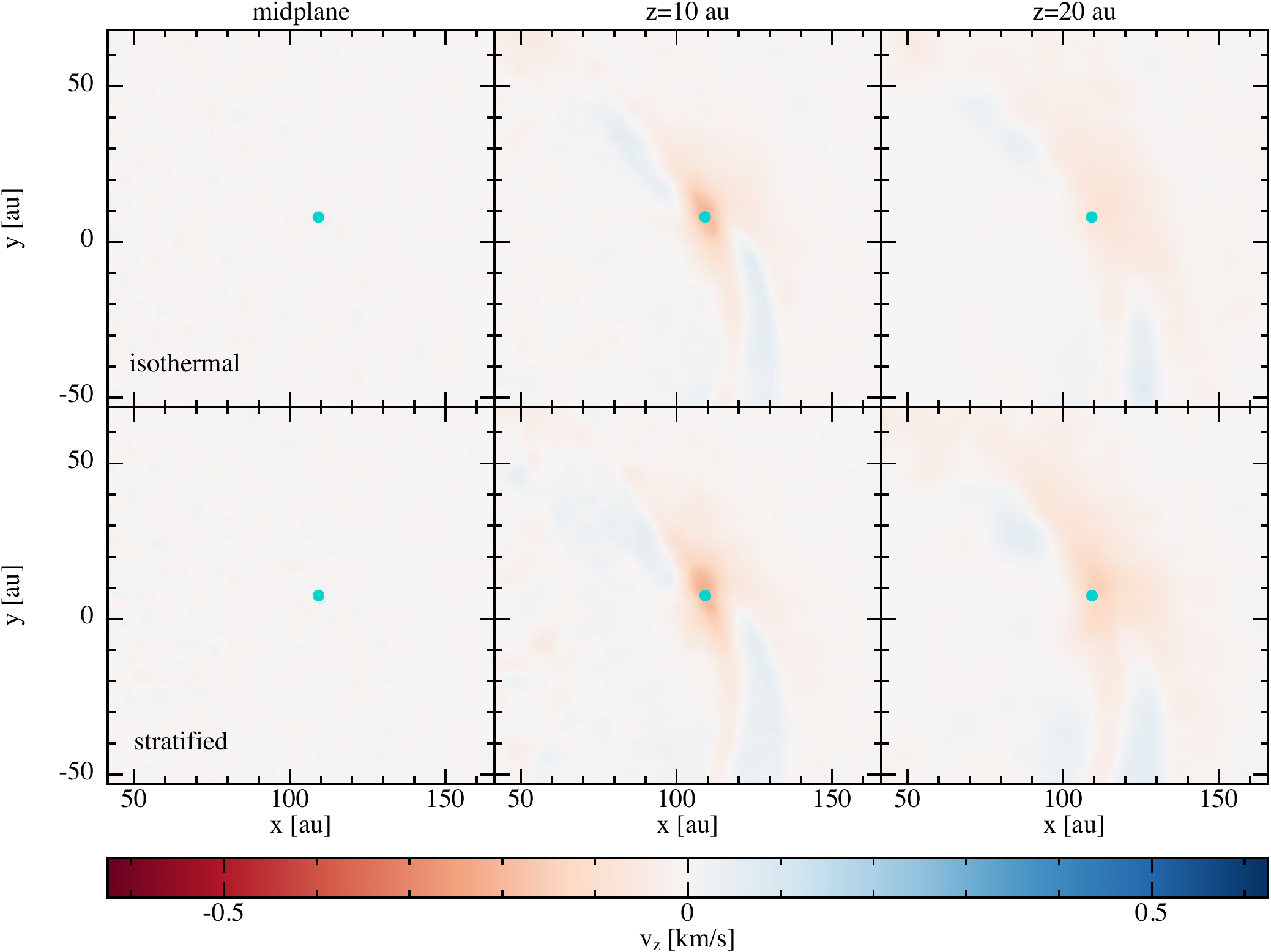}
\caption{Vertical velocity perturbation induced by a 2M$_\mathrm{Jup}$ planet in a vertically isothermal calculation (top) compared to a calculation with stratified vertical temperature (bottom). We show slices taken in the midplane (left) and at heights of 10 and 20 au above the midplane (middle and right, respectively). \label{sup_fig:vz}}
\end{figure}

\begin{figure}
  \centering
  \includegraphics[width=0.7\linewidth,angle=0]{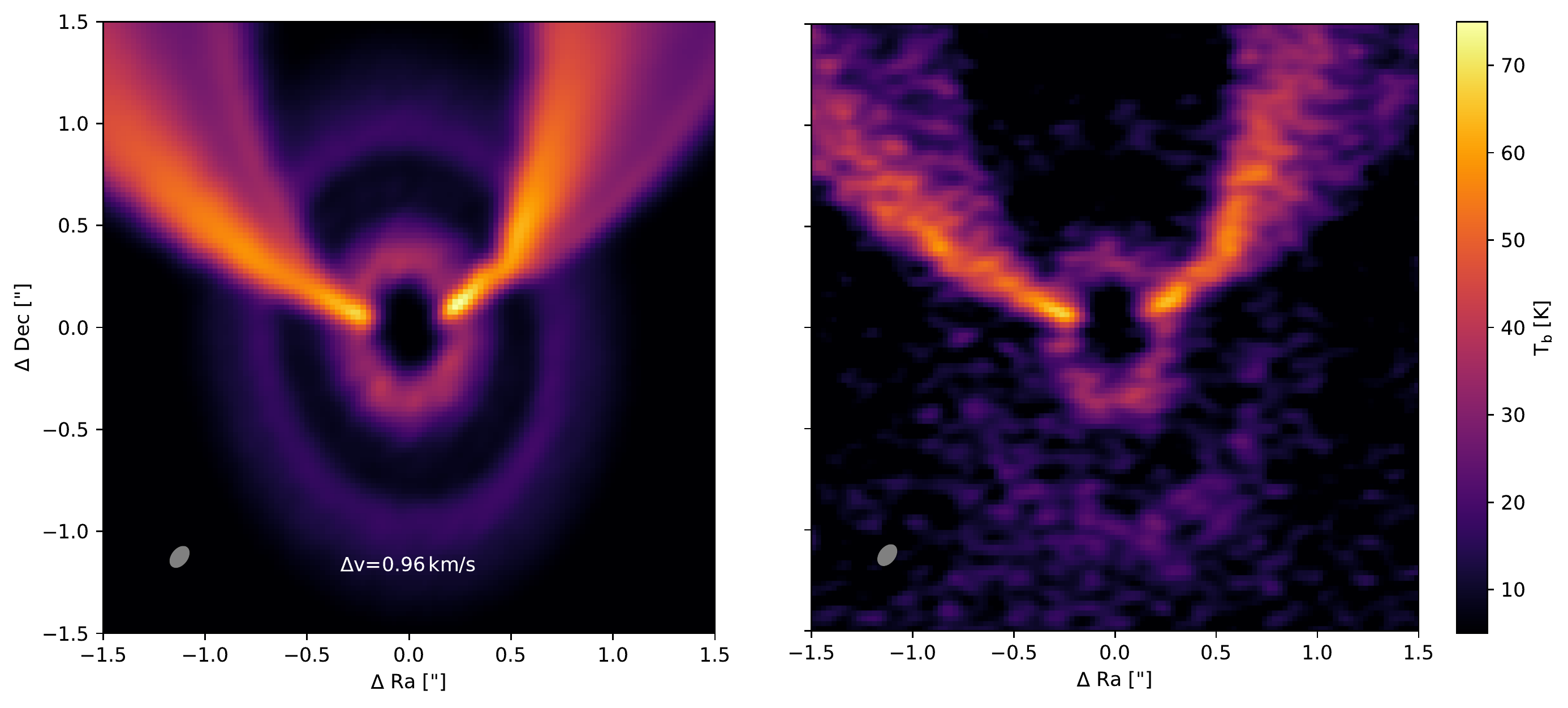}
  \caption{Comparison of the same channel map of the 2\,M$_\mathrm{jup}$ model with convolution by a Gaussian beam (left) or by sampling the synthetic visibilities at the same \emph{uv} points as the data, and re-imaging using the same parameters as the data (right). \label{sup_fig:test_conv}}
\end{figure}

\end{document}